# Heat loads to the wall: the thermal conduction approach

M Kovari

CCFE, Culham Science Centre, Abingdon, Oxon, OX14 3DB, UK.
Tel.: +44 (0)1235-46-6427. E-mail address: michael.kovari@ukaea.uk

## Abstract

Calculations of the heat flux carried by plasma to the wall of a magnetic fusion machine often assume that power flows only along the field lines, but this cannot be true in general. Instead, we treat the plasma as an anisotropic non-linear thermally conducting medium. The model is physically relevant if parallel and cross-field transport are driven at least in part by temperature gradients, which means they are affected by the proximity of material surfaces. The model generates a familiar asymmetrical power distribution on the divertor target, and divergent heat flux on sharp edges, as described previously. The power landing on a wall that is parallel to the field (identically zero in the conventional model) is a substantial fraction of that landing on the target. The flux profile across the scrape-off layer (SOL) deviates strongly from an exponential, because the parallel heat conduction is very large near the separatrix, giving a much steeper fall-off in that region.

When a portion of the wall protrudes into the SOL at a grazing angle, the power striking the inclined surface is several times the value derived by integrating the parallel heat flux calculated in the absence of the inclined surface. We have also calculated the degree to which the shadowing effect of a first wall component on other parts of the wall falls off with distance – something that cannot even be estimated using the field-line following technique.

## 1.      Introduction

Calculations of the heat flux carried by plasma to the wall of a magnetic fusion machine often assume that heat flows only along the field lines. This cannot be true in general, as the very existence of the scrape-off layer (SOL) depends on cross-field transport. Instead, Goldston (1) has proposed treating the plasma as an anisotropic non-linear thermally conducting stationary medium, capable of conducting heat to a surface in contact with it. This is the simplest model capable of describing the competition between parallel and cross-field transport. Note that this model is distinct from the "particle funnelling" effect proposed by (2), which is based on plasma flow.

We have explored the consequences of this model for the SOL of a divertor tokamak, in the following four situations:
   a) A wall that is perfectly conformal to the field lines
   b) A panel with exposed radial edges on both sides
   c) A panel protruding from the wall at a small angle to the field
   d) An exposed radial edge, shadowed by an upstream panel.

We have also examined a limiter plasma configuration, where there is experimental evidence that field line mapping of exponential SOL power profiles does not give correct results (3).

The heat fluxes given below are decomposed into components parallel or perpendicular to the field. Those in the parallel direction can be described as "parallel heat flux", $q_\parallel$, but it is important to emphasise that this flux depends on the presence of material surfaces, and cannot be considered as a fixed output from the confined plasma.

## 2.      The model

We have modelled a flux tube in the SOL, effectively straightened out into a rectangle. The $x$ axis is locally normal to the flux surface, and represents the distance from separatrix. The $y$ axis represents the distance measured along the field line from the plasma midplane (the connection length). The $z$ coordinate is normal to the other two, lying in the flux surface, and represents the "width" of the flux tube. The model is two-dimensional, so all quantities are independent of $z$. Note that this does not correctly describe the toroidal symmetry, as $z$ is not toroidal. Figure

1 illustrates a small portion of a flux tube. If the pitch angle of the field line at the midplane (angle between field line and equator) is α, the total width in the z direction of all the flux tubes is

$$2\pi(R_0 + a)\tan(\alpha),$$

where $R_0$ is the plasma major radius, and a is the minor radius. For a reactor-sized machine where $R = 9$ m, $a = 3$ m, $\tan(\alpha) = 0.25$, this width is 4.7 m.

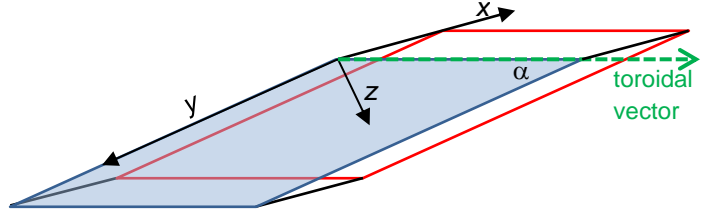

Figure 1. A segment of a flux tube. The front face (shaded) represents a portion of the separatrix. x is the minor radius vector normal to the separatrix, y is parallel to the magnetic field. The pitch angle of the field line is α.

We have used a standard Spitzer-type parallel thermal conductivity for deuterium (4):

$$\kappa_\parallel = \kappa_{0e} T^{\frac{5}{2}} \quad where \quad \kappa_{0e} \approx 2000 \frac{W}{m \cdot eV^{7/2}}.$$

There is no general expression for a multi-species plasma, but a form that may be approximately correct when impurities are present is proposed by (5):

$$\kappa_{0e} \approx \frac{8788}{Z_{eff}}\left(\frac{Z_{eff} + 0.21}{Z_{eff} + 4.2}\right)\frac{W}{m \cdot eV^{7/2}},$$

where $Z_{eff}$ is the effective ion charge. The cross-field thermal conductivity is not known, but we have followed Goldston in using a Bohm-like dependence so the conductivity is proportional to temperature:

$$\kappa_\perp = kT \quad where \quad k = 1.6 \frac{W}{m \cdot eV} \quad unless\ stated.$$

The coefficient *k* conveniently summarises the key unknown in this model, and it is easy to show that uncertainty in *k* is equivalent to uncertainty in the cross-field length scale. The effects of varying the constant of proportionality on the SOL width are considered in section 3.1. Plasma flow, density variations, neutrals, radiation and other mechanisms are ignored. The model is strongly non-linear because the parallel conductivity varies strongly with temperature.

The Ansys finite element code was used, with the Workbench interface. For convenience all dimensions in the y direction were reduced by a factor of 1000. The required scalings are given in Table . Ansys occasionally fails to find a solution when the non-linearity is too extreme, so to limit this, the temperature boundary condition at the target and wall was set to 3 eV unless otherwise stated. The calculation for a single geometry takes about 1 minute on single-core PC.

A sheath boundary condition would be more physically relevant. A single comparison model was run incorporating this in the form of a temperature-dependent thermal contact resistance, with a fixed plasma density. This condition was applied only at the target, as the sheath condition for the wall, parallel to the field, is not known.

Table 1. Scaling factors (true value / scaled value). *The gradients of the heat flux appear in the heat conduction equation and must have the same scaling factor.

| Quantity or operator | Symbol | Scaling factor |
|---|---|---|
| Cross-field dimensions | $x, z$ | 1 |
| Parallel dimension | $y$ | $f = 1000$ |
| Cross-field gradient | $\frac{\partial}{\partial x}$ | 1 |
| Parallel gradient | $\frac{\partial}{\partial y}$ | $\frac{1}{f}$ |
| Gradient of heat flux* | $\frac{\partial}{\partial x}\left(\kappa_{xx}\frac{\partial T}{\partial x}\right)$ | 1 |
| Gradient of heat flux* | $\frac{\partial}{\partial y}\left(\kappa_{yy}\frac{\partial T}{\partial y}\right)$ | 1 |
| Cross-field thermal conductivity | $\kappa_{xx}\ \kappa_{zz}$ | 1 |
| Parallel thermal conductivity | $\kappa_{yy}$ | $f^2$ |
| Cross-field heat flux | $\kappa_{xx}\frac{\partial T}{\partial x}$ | 1 |
| Parallel heat flux | $\kappa_{yy}\frac{\partial T}{\partial y}$ | $f$ |
| Cross-field power | $\kappa_{xx}\frac{\partial T}{\partial x}\Delta y\Delta z$ | $f$ |
| Parallel power | $\kappa_{yy}\frac{\partial T}{\partial y}\Delta x\Delta z$ | $f$ |

## 2.1. Divertor configuration

In a divertor configuration the cross-section of the flux tube varies strongly as it approaches the X-point: the radial separation of flux surfaces expands, while the width in the *z* direction contracts. Moreover, the connection length tends to infinity as the distance from the separatrix tends to zero. In our preliminary model neither of these effects are taken into account. Instead, the flux tube is regarded as a simple block, with a constant extent in each dimension. Effectively the flux tube immediately adjacent to the separatrix, with very large connection length, is ignored because very little heat can be conducted along this flux tube to the target. Figure 2 shows the layout of the model. It is generally believed that the power entering the SOL is concentrated near the outboard midplane where cross-field transport is likely to be greatest, and we have represented this by passing power only through the half of the separatrix nearest the midplane. Note that while a conventional divertor target is inclined relative to the field lines, this cannot be reproduced in this 2D model.

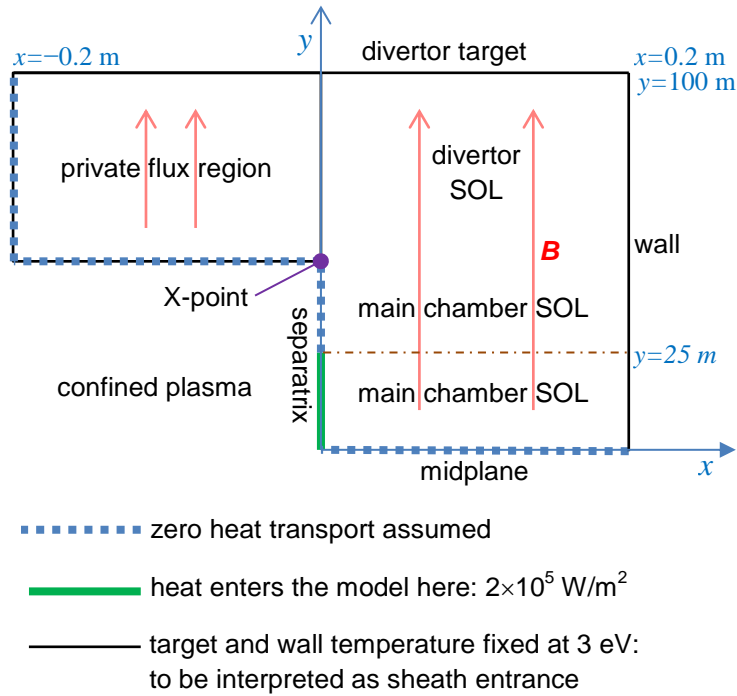

Figure 2. The divertor model. The extent in the z direction is set to 1 m without loss of generality. The connection length from midplane to target (100 m) is taken to represent a typical DEMO design. The heat input is 25 m × 2×10$^5$ W/m$^2$ = 5×10$^6$ W/m, or 2.35×10$^7$ W in total.

## 2.2. Limiter configuration

A limiter plasma is shown in Figure 3, with the 2D model on the right.

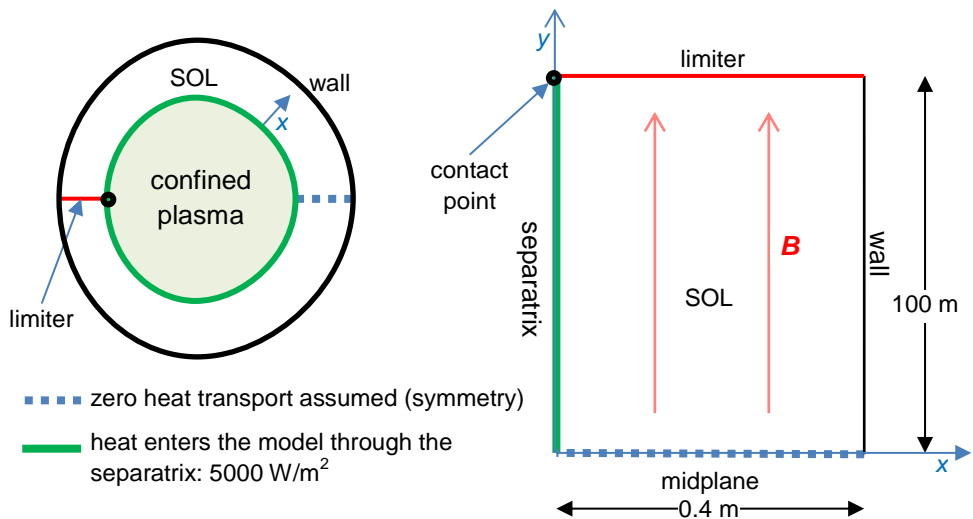

Figure 3. Limiter model. Left: real world geometry in poloidal plane. Right: Rectilinear model. The extent in the z direction is set to 1 m without loss of generality. The heat input per unit extent in z is 100 m × 5000 W/m$^2$ = 5×10$^5$ W/m.

## 3. Results: Divertor configuration

### 3.1. Conformal wall

The wall is as shown in Figure 4, conformal to the field lines. In the conventional approach the heat flux on the wall is exactly zero. Figure 1 shows the temperature distribution in the scaled model.

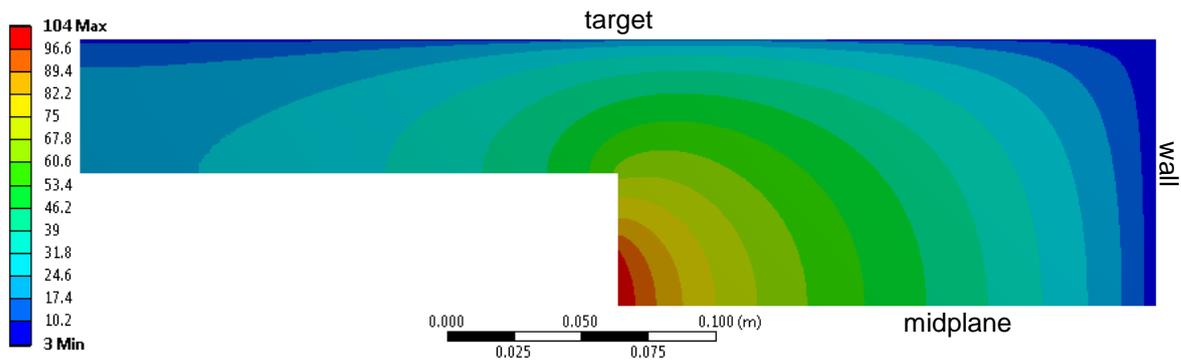

Figure 1. The temperature distribution in the SOL (eV) with a conformal wall. See Figure 2 for an explanation of the layout. Scaled model.

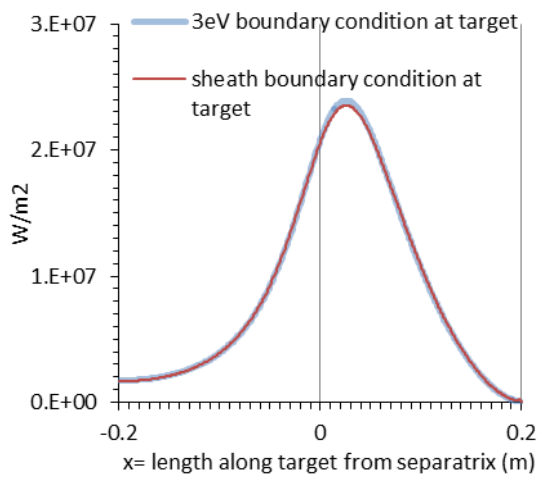

Figure 5. Parallel heat flux to the target, conformal wall. The left hand half represents the private flux region. Note that the boundary conditions on the two ends of the target are different.

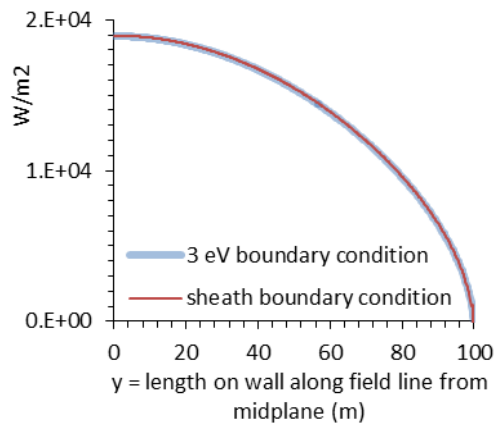

Figure 6. Cross-field heat flux to the conformal wall

Figure 5 shows the familiar asymmetrical power distribution on the target. Figure 6 shows that the power density on the wall is indeed small in comparison. However, the total power on the wall per unit interval of $z$ is substantial: 1.4 MW/m, compared to 3.6 MW/m on the target. In both figures the results using the 3 eV boundary condition used throughout this paper and those using the sheath condition at the target (section 2 above) are almost identical.

The heat flux profile across the SOL (Figure 7) deviates strongly from an exponential. The exponential fit shown gives a fall-off length $\lambda_q = 26$ mm, but near the separatrix the parallel heat conduction is very large, giving a much steeper fall-off. Near the wall the parallel heat flux drops to zero as one approaches the constant temperature wall.

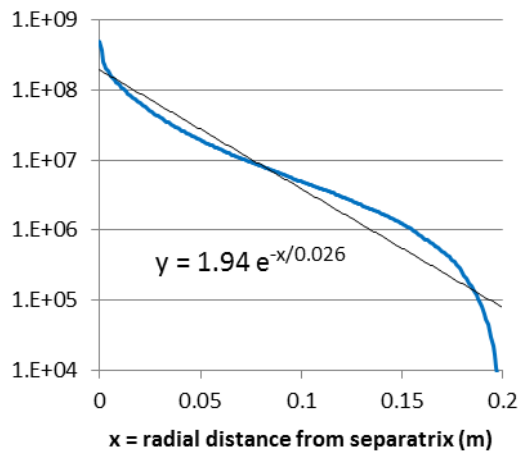

Figure 7. Parallel heat flux (W/m$^2$), at y = 25 m (shown by the dash-dot line in Figure ).

Figure 8 shows the fraction of the total power that lands on the wall, as a function of the gap between the separatrix and the wall, for various values of the cross-field transport coefficient.

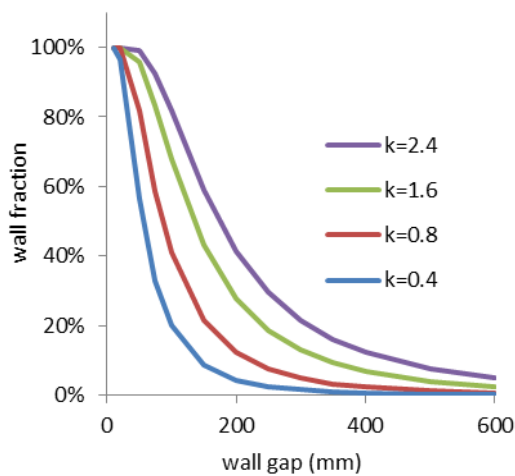

Figure 8. Power fraction landing on the wall as a function of the gap between the separatrix and the wall, for various values of the cross-field transport coefficient k (W/m/eV).

## 3.2. Rectangular panel

Figure 9 shows the effect on the local temperature distribution when a rectangular panel protrudes from the wall.

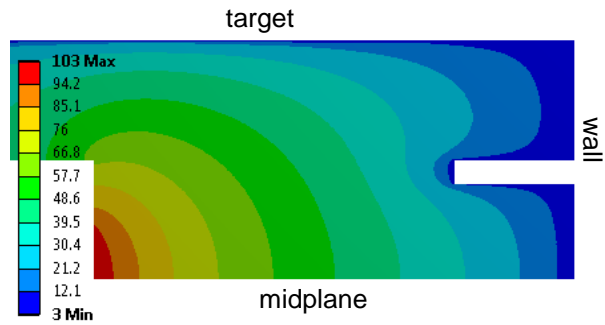

Figure 9. Temperature distribution (eV) when a rectangular panel protrudes from the wall, 10 m long x 50 mm deep radially.  Scaled model.  Part of the private flux region is not shown.

The mesh in the vicinity of the corner of the panel is shown in Figure 10.  The discretisation does not introduce significant errors in the power balance (Table 2).

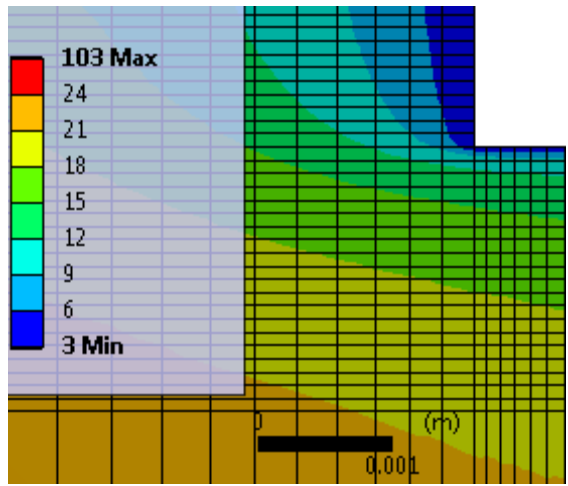

Figure 10.  Temperature (eV) in the vicinity of the corner of the rectangular panel, with mesh elements.  The ruler shows scaled dimensions.

Table 2.  Power balance for unit length in the *z* direction, model with rectangular panel

| Power input | 5 MW/m |
| --- | --- |
| Power on target | 3.3165 MW/m |
| Power on wall | 0.78069 MW/m |
| Power on panel | 0.90091 MW/m |
| Output power / input power | 0.9996 |

Figure 11 illustrates the divergence that occurs at the sharp corners of the panel, described by (1).

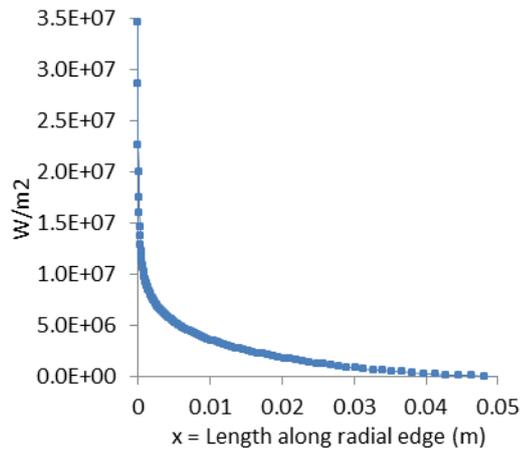

Figure 11. Parallel heat flux on the upstream edge of the rectangular panel (see Figure ). The last point has a large discretisation error and is not shown.

### 3.3. Sloping panel

Figure 12 shows the effect on the local temperature distribution when an inclined panel is included at a small angle to the field.

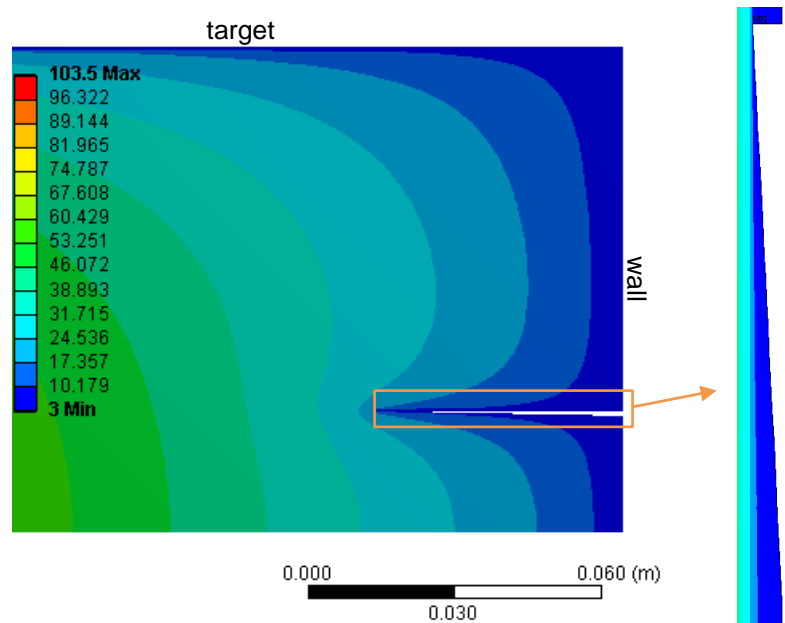

Figure 12. Temperature distribution (eV) near a sloping panel, 1 m long x 50 mm deep radially. Left: scaled model. Right: detail of panel with actual proportions.

Figure 13 gives the total power landing striking the inclined face of the panel, and shows that the estimate obtained by integrating the parallel heat flux $q_\parallel$ derived in the *absence* of the panel does *not* give the correct answer. The proximity of the cold panel increases the temperature gradient in the SOL.

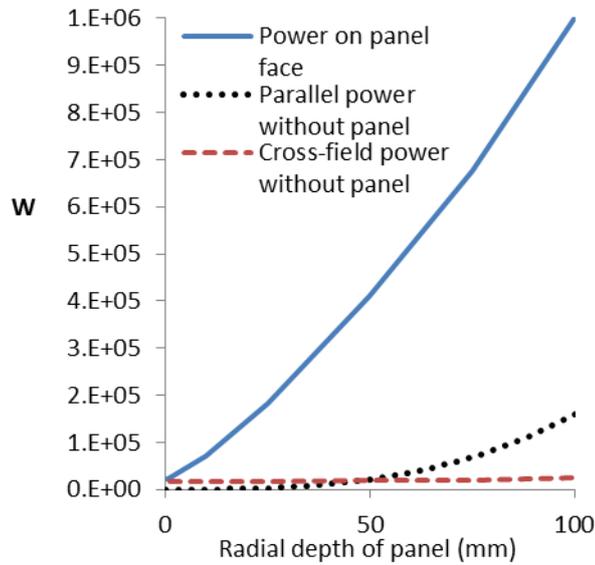

Figure 13. Power on the inclined face of the panel (W). The dotted and dashed curves shows the prediction obtained by integrating the parallel and cross-field heat flux derived for the conformal wall case.

### 3.4. Shadowing of an exposed radial edge

An exposed radial edge can be shadowed by another component (Figure 14), but only if it is sufficiently nearby. Figure 15 shows the degree of shadowing for a range of values of the distance between the two components. This cannot even be estimated using the field-line following technique. This effect is connected to the "collection length" introduced by Stangeby (6) – the length over which the disturbance to the plasma extends, before cross-field transport exceeds the particle and power flux that can be transported through the sheath in front of the object. Of course our model takes into account neither the sheath nor the "short-circuit" effect that occurs when flux tubes become strongly sheared.

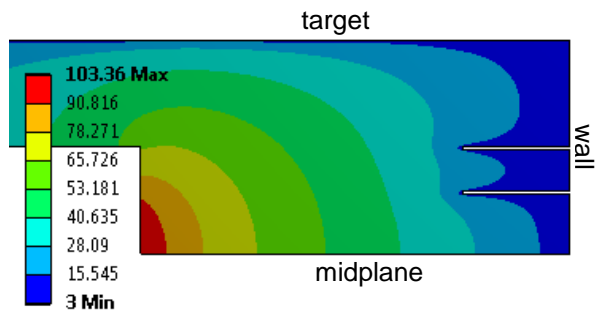

Figure 14. The temperature distribution in the SOL (eV) with two panels shadowing each other. Scaled model. Part of the private flux region has been omitted for clarity.

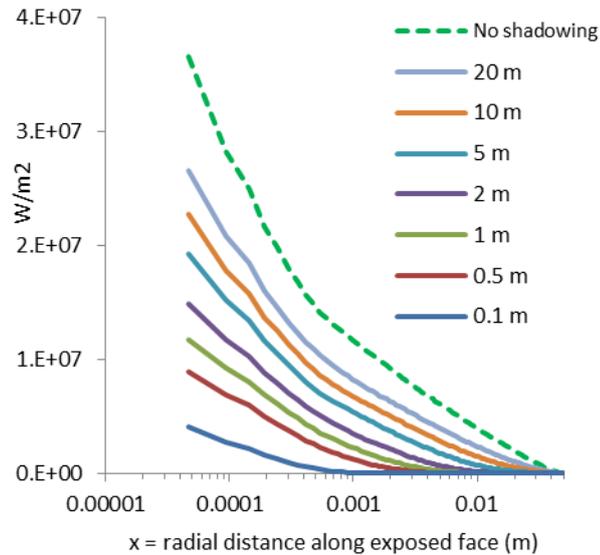

Figure 15. Parallel heat flux on the upstream edge of the downstream panel (see Figure ), for various separations between the panels.

## 4.     Results: Limiter configuration

Arnoux *et al* (3) found that the 2D thermal conduction model was "not inconsistent" with JET limiter results, but felt that it was nevertheless unphysical, as it required a large parallel temperature gradient. To address this we have run the model for a range of values of limiter temperature. (The wall temperature is set to the same value to avoid unphysical heat flows.) Of course the physical wall itself is cold (temperature << 1 eV), but there will certainly be a sheath, and there may also be a narrow zone next to it with high neutral density and a large temperature gradient, reducing the temperature differential in the main part of the SOL. The specified "limiter temperature" should be taken as the temperature at the upstream edge of this zone. Figure 16 shows that when the limiter is raised from 3 to 30 eV, the temperature gradient in the SOL drops accordingly, but the heat flux still tends to infinity in a qualitatively similar way as one approaches the contact point. (Note that the horizontal axis is logarithmic, and the heat flux profile shows an approximately logarithmic divergence – not to be confused with an exponential decay!) We feel this model may in fact be appropriate for a limiter.

Note that the profiles shown here are for a radial limiter in the midplane. We would argue that it is not possible to directly convert this result to an inner wall limiter, tangential to a flux surface, because the flux profile depends on the material surfaces. Unfortunately the 2D nature of the model makes it impossible to incorporate the relevant angles directly.

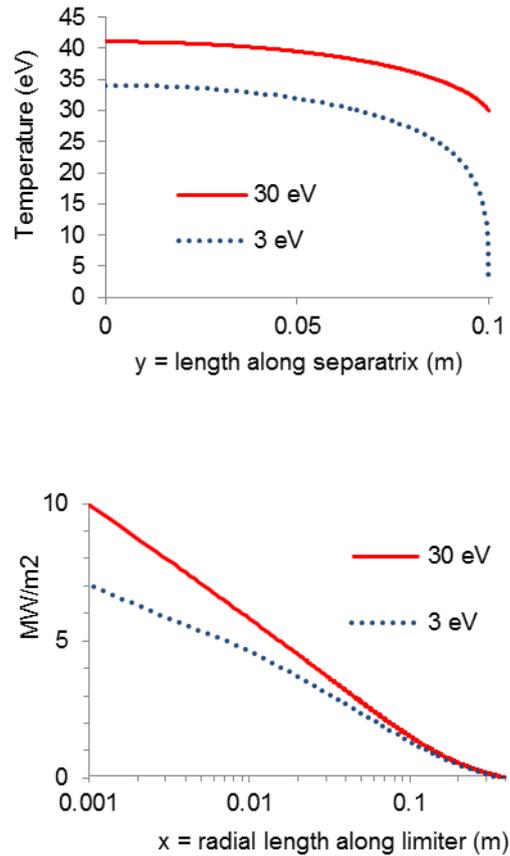

Figure 16. Inner midplane limiter configuration, temperature of limiter and wall set to 3 eV and 30 eV. Top: temperature along the separatrix from the midplane. Bottom: Parallel heat flux profile on the limiter. The contact point is at $x = 0$. See Figure 3 for a sketch.

## 5. Discussion

The thermal conduction model used here is physically relevant if both parallel and cross-field transport are driven at least in part by temperature gradients, which means that a material surface nearby will increase heat flux. This leads to effects that cannot be predicted by field-line following. Compared to state-of-the art codes like SOLPS, which takes 2 months to run, even when the effective CPU speed exceeds 10 Gflops (7), this approach gives scope for exploration of different geometries and parameters.

A sheath boundary condition would be more physically relevant. In a preliminary comparison the results using the 3 eV boundary condition used throughout this paper and those using a sheath condition at the target are almost identical. Another possible improvement would be to use the poloidal coordinate instead of the connection length used here. This would allow real axisymmetric limiter geometry to be included.

## Acknowledgments

I would like to thank B. Sieglin, E. Surrey and R. Goldston for valuable advice. This work was part-funded by the RCUK Energy Programme [grant number EP/I501045] and has received funding from the Euratom research and training programme 2014-2018 under grant agreement No 633053, within the framework of the EUROfusion Consortium. To obtain further information on the models underlying this paper please contact the authors, or PublicationsManager@ccfe.ac.uk. The views and opinions expressed herein do not necessarily reflect those of the European Commission.

# Appendix: Supplementary data

Ansys reports have been uploaded.